\documentclass{article}

\usepackage{arxiv}

\usepackage[utf8]{inputenc} 
\usepackage[T1]{fontenc}    
\usepackage[hidelinks]{hyperref}
\usepackage{graphicx}
\usepackage{caption}
\usepackage{subcaption}
\usepackage{url}            
\usepackage{booktabs}       
\usepackage{amsfonts}       
\usepackage{nicefrac}       
\usepackage{microtype}      
\usepackage{lipsum}		
\usepackage[round]{natbib}
\usepackage{doi}
\usepackage{amsthm}
\usepackage{amsmath}
\usepackage{mathabx}
\usepackage{diagbox}
\usepackage{stackengine}
\usepackage{algorithm}
\usepackage{algpseudocode}
 \usepackage{setspace}
\let\Algorithm\algorithm
\renewcommand\algorithm[1][]{\Algorithm[#1]\setstretch{1.2}}

\usepackage{array}
\usepackage{eqparbox}

\newtheorem{thm}{Theorem}

\title{General Form Moment-based Estimator of Weibull, Gamma, and Log-normal Distributions}

\date{} 					

\author{ 
\href{https://orcid.org/0000-0003-2002-984X}{\includegraphics[scale=0.06]{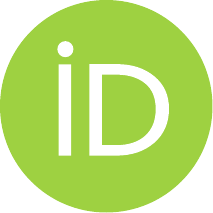}\hspace{1mm}Kang Liu}\thanks{Personal website: https://cruyffio.github.io/mysite/} \\
Independent Researcher\\
\texttt{liukangk11@gmail.com} \\
}

\renewcommand{\undertitle}{}

\hypersetup{
pdftitle={A template for the arxiv style},
pdfsubject={q-bio.NC, q-bio.QM},
pdfauthor={David S.~Hippocampus, Elias D.~Striatum},
pdfkeywords={First keyword, Second keyword, More},
}

\begin{document}
\maketitle

\vspace{-20pt}

\begin{abstract}
This paper presents a unified and novel estimation framework for the Weibull, Gamma, and Log-normal distributions based on arbitrary-order moment pairs. Traditional estimation techniques, such as Maximum Likelihood Estimation (MLE) and the classical Method of Moments (MoM), are often restricted to fixed-order moment inputs and may require specific distributional assumptions or optimization procedures. In contrast, our general-form moment-based estimator allows the use of any two empirical moments—such as mean and variance, or higher-order combinations—to compute the underlying distribution parameters. For each distribution, we develop provably convergent numerical algorithms that guarantee unique solutions within a bounded parameter space and provide estimates within a user-defined error tolerance. The proposed framework generalizes existing estimation methods and offers greater flexibility and robustness for statistical modeling in diverse application domains. This is, to our knowledge, the first work that formalizes such a general estimation structure and provides theoretical guarantees across these three foundational distributions.

\end{abstract}


\section{Introduction}
The Weibull, Gamma, and Log-normal distributions are three widely used continuous probability distributions that play a crucial role in statistical modeling \citep{freedman2009statistical,kroese2014statistical,chambers2017statistical}, reliability engineering \citep{zio2009reliability,kapur2014reliability,xu2021machine}, queueing theory \citep{giambene2005queuing,ng2008queueing,gross2011fundamentals}, production system engineering \citep{alavian2018alpha,alavian2019alpha,alavian2020alpha,eun2022production,alavian2022alpha,liu2021alpha}, robotics \citep{cheng2011reliability,abiri2017tensile,li2019imprecise,liu2019vehicle}, and even machine learning \citep{von2022knowledge,liu2024setcse}. Each distribution is characterized by its flexibility in modeling skewed data and its relevance in domains where understanding variability, failure rates, or time-to-event behavior is essential.

In probability theory and statistics \citep{blom2012probability,rohatgi2015introduction}, the moments of a random variable provide a quantitative description of its distribution's shape and characteristics. Formally, the $k$th moment of a random variable $X$ is defined as the expected value of its $k$th power, $E(X^k)$, provided the expectation exists. Moments capture important distributional features such as location, spread, skewness, and kurtosis, and are foundational in both theoretical and applied statistics.

In this paper, we introduce a novel general-form moment estimation framework for the Weibull, Gamma, and Log-normal distributions. Unlike traditional estimation methods such as Maximum Likelihood Estimation (MLE) or the classical Method of Moments (MoM), our approach allows the use of any two statistical moments of arbitrary order—for example, mean and variance, variance and skewness, or higher-order combinations—to estimate distribution parameters. This flexibility makes the framework adaptable to a wide range of practical scenarios where different summary statistics are more accessible or meaningful.

We develop tailored algorithms for each distribution and prove that the algorithms are guaranteed to have a unique solution within a compact, bounded parameter space. Furthermore, we show that the estimation procedure can be carried out to within a user-specified error tolerance, ensuring computational reliability. To the best of our knowledge, this is the first work to present such a unified and theoretically grounded estimator that generalizes moment-based inference across these three important distributions. This framework not only encompasses traditional estimators as special cases but also opens up new possibilities for robust and interpretable statistical modeling in diverse applied domains.

\section{Background}

Estimating the parameters of a probability distribution from observed data is a fundamental task in statistical inference, and it directly influences the accuracy and interpretability of probabilistic models. For the Weibull, Gamma, and Log-normal distributions, several well-established estimation techniques are commonly employed, each with its own strengths depending on sample size, data characteristics, and the target application.

One of the most widely used approaches is Maximum Likelihood Estimation (MLE) \citep{pan2002maximum,myung2003tutorial,chernoff2011use,groeneboom2012information}, which identifies parameter values that maximize the likelihood function given the observed data. MLE is asymptotically efficient and provides consistent parameter estimates, but may require iterative numerical optimization, especially for the Weibull and Gamma distributions due to the absence of closed-form solutions.

Another frequently used method is the Method of Moments (MoM) \citep{erickson2002two,hansen2010generalized,forbes2011statistical,zsohar2012short,abbaszadehpeivasti2023method}, which equates the theoretical moments of the distribution to the corresponding sample moments. This technique is generally simpler to compute and provides reasonable estimates, particularly when sample sizes are moderate. For example, in the case of the Gamma distribution, MoM leads to explicit expressions for the shape and scale parameters using the sample mean and variance.

In some applications, Bayesian methods are also adopted \citep{zong2000bayesian,larranaga2001estimation,kaminskiy2005simple,glickman2007basic,kruschke2013bayesian,li2014application,maritz2018empirical}, particularly when prior information is available or when robust estimates are required under uncertainty. In the context of reliability analysis or survival modeling, regression-based techniques and censored data handling may also be integrated into the estimation process.

\section{Weibull, Gamma and Log-normal Distribution and their Moments}

\begin{figure}
\centering

\begin{subfigure}{0.45\textwidth}
  \centering
  \includegraphics[width=0.8\linewidth]{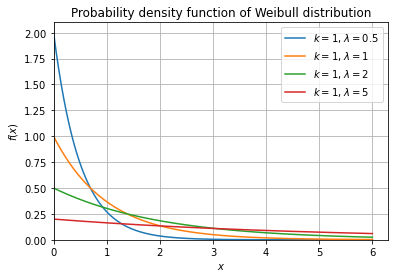}
  \caption{$k=1$}
  \label{fig:weibull_pdf1}
\end{subfigure}%
\begin{subfigure}{0.45\textwidth}
  \centering
  \includegraphics[width=0.8\linewidth]{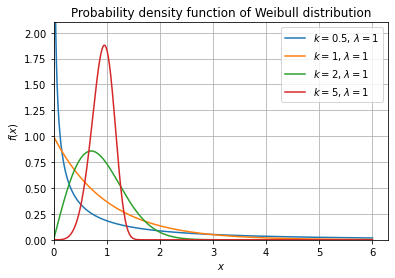}
  \caption{$\lambda=1$}
  \label{fig:weibull_pdf2}
\end{subfigure}

\caption{Probability density function of Weibull distribution.}
\label{fig:weibull_pdf}
\end{figure}

The probability density function of a Weibull random variable is given by the following formula:
\begin{equation}\label{eq:weibull_pdf}
f(x; k, \lambda)= \begin{cases}
\frac{k}{\lambda} \left(\frac{x}{\lambda}\right)^{k-1}  e^{-\left(\frac{x}{\lambda}\right)^k} &\text{ for } x\geq 0 \\
0 &\text{ for } x<0
\end{cases}
\; \text{ ,}
\end{equation}
where $k > 0$ is the shape parameter and $\lambda > 0$ is the scale parameter of the distribution. The probability density function is shown in Figure \ref{fig:weibull_pdf} for different values of $k$ and $\lambda$. In addition, the $i$-th moment of Weibull distribution is:
\begin{equation}
E(X^i) = \lambda^i \Gamma\left(1 + \frac{i}{k}\right) \text{ .}
\end{equation}

\begin{figure}
\centering

\begin{subfigure}{0.45\textwidth}
  \centering
  \includegraphics[width=0.8\linewidth]{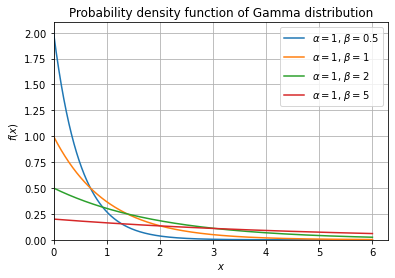}
  \caption{$\alpha=1$}
  \label{fig:gamma_pdf1}
\end{subfigure}%
\begin{subfigure}{0.45\textwidth}
  \centering
  \includegraphics[width=0.8\linewidth]{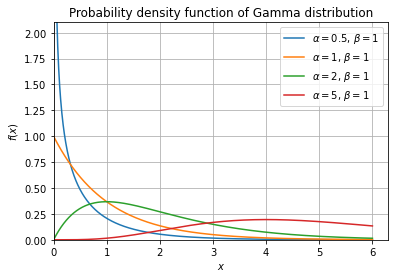}
  \caption{$\beta=1$}
  \label{fig:gamma_pdf2}
\end{subfigure}

\caption{Probability density function of Gamma distribution.}
\label{fig:gamma_pdf}
\end{figure}

The probability density function of a Gamma random variable is given by the following formula:
\begin{equation}\label{eq:gamma_pdf}
f(x; \alpha, \beta)= \begin{cases}
\frac{1}{\beta^{\alpha} \Gamma(\alpha)} x^{\alpha-1} e^{\frac{-x}{\beta}} &\text{ for } x \geq 0
 \\
0 &\text{ for } x<0
\end{cases}
\; \text{ ,}
\end{equation}
where $\alpha > 0$ is the shape parameter and $\beta > 0$ is the scale parameter of the distribution. The probability density function is shown in Figure \ref{fig:gamma_pdf} for different values of $\sigma$ and $\mu$. In addition, the $i$-th moment of Gamma distribution is:
\begin{equation}\label{eq:moment_gamma}
E\left(X^i\right) = \frac{\beta^i \Gamma(i+\alpha)}{\Gamma(\alpha)} \text{ .}
\end{equation}

\begin{figure}
\centering

\begin{subfigure}{0.45\textwidth}
  \centering
  \includegraphics[width=0.8\linewidth]{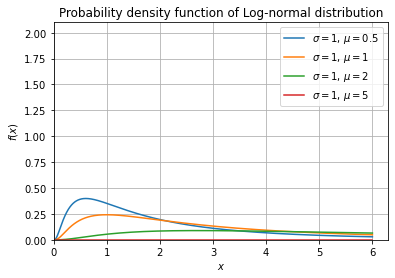}
  \caption{$\sigma=1$}
  \label{fig:lognor_pdf1}
\end{subfigure}%
\begin{subfigure}{0.45\textwidth}
  \centering
  \includegraphics[width=0.8\linewidth]{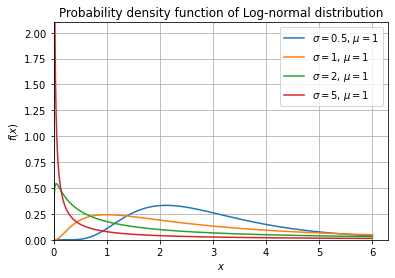}
  \caption{$\mu=1$}
  \label{fig:lognor_pdf2}
\end{subfigure}

\caption{Probability density function of Log-normal distribution.}
\label{fig:lognor_pdf}
\end{figure}

The probability density function of a Log-normal random variable is given by the following formula:
\begin{equation}\label{eq:lognor_pdf}
f(x; \mu, \sigma)= \begin{cases}
\frac{1}{\sqrt{2\pi}\sigma x} e^{-\frac{(\ln{x} - \mu)^2}{2\sigma^2}} &\text{ for } x \geq 0
 \\
0 &\text{ for } x<0
\end{cases}
\; \text{ ,}
\end{equation}
where $\mu \in \mathbb{R}$ denotes the logarithm of location, and $\sigma > 0$ denotes logarithm of scale. The probability density function is shown in Figure \ref{fig:lognor_pdf} for different values of $\sigma$ and $\mu$. In addition, the $i$-th moment of Log-normal distribution is:
\begin{equation}
E\left(X^i\right) = e^{\mu i + \frac{1}{2} \sigma^2 i^2 } \text{ .}
\end{equation}

\section{General Form Moment-based Estimator of Weibull Distribution}

\begin{thm}\label{thm:weibull}
For a Weibull random variable with the specified moments of orders $n$ and $m$, i.e., $E(X^n)$ and $E(X^m)$, where $n>m$, its parameters $\lambda$ and $k$ can be solved numerically by Algorithm \ref{alg:weibull}.
\end{thm}

\begin{algorithm}[H]
\caption{Solving Weibull parameters numerically given a pair of moments}\label{alg:weibull}
\hspace*{\algorithmicindent} \textbf{Input:} $n$, $m$, $E(X^n)$, $E(X^m)$, tolerance $\delta$\\
\hspace*{\algorithmicindent} \textbf{Output: $k$, $\lambda$}
\begin{algorithmic}[1]
\State $k_{\text{min}} \gets \widecheck{K}$, $k_{\text{max}} \gets \widehat{K}$ \Comment{choose $\widecheck{K}$ and $\widehat{K}$ such that the solution of $k$ is within $[\widecheck{K}, \widehat{K}]$}
\State $r_W \gets {E^{m}(X^n)}/{E^{n}(X^m)}$
\While{$k_{\text{max}} - k_{\text{min}} > \delta$}
\State $k_{\text{mid}} \gets {(k_{\text{max}} + k_{\text{min}})}/{2}$
\If{${\Gamma^{m}(1 + {n}/{k_{\text{mid}}})}/{\Gamma^{n}(1 + {m}/{k_{\text{mid}}})} > r_W$}
    \State $k_{\text{min}} \gets k_{\text{mid}}$
\Else
    \State $k_{\text{max}} \gets k_{\text{mid}}$
\EndIf
\EndWhile
\State $k \gets {(k_{\text{max}} + k_{\text{min}})}/{2}$
\State $\lambda \gets \sqrt[m]{E(X^m)/\Gamma(1 + m/k)}$
\end{algorithmic}
\end{algorithm}

\begin{proof}
Denote the ratio of $E^{m}(X^n)$ and $E^{n}(X^m)$ as $r_W$, we have
\begin{equation}
    r_W = R_{W}(x) = \frac{E(X^n)^m}{E(X^m)^n} = \frac{\lambda^{mn}\Gamma^m\left(1 + \frac{n}{k}\right)}{\lambda^{mn}\Gamma^n\left(1 + \frac{m}{k}\right)} = \frac{\Gamma^m\left(1+\frac{n}{k}\right)}{\Gamma^n\left(1+\frac{m}{k}\right)}.
\end{equation}
In Section 4.1 of \cite{liu2025moment}, it is shown that $R_W(k)$ is monotonically decreasing with respect to $k$.

By the Intermediate Value Theorem (continuity), since $r_W \in [R_W(k_{\text{min}}), R_W(k_{\text{max}})]$ there exists $k^{*} \in [k_{\text{min}}, k_{\text{max}}]$ such that $r_W = R_W(k^{*})$.

At each step of Algorithm \ref{alg:weibull}, the length of the interval halves. Thus, after $h$ steps, $k^h_{\text{max}} - k^h_{\text{min}} = \frac{1}{2^h}(k_{\text{max}} - k_{\text{min}})$. In other words, the length of the interval shrinks exponentially fast.

Since the intervals $[k^h_{\text{min}}, k^h_{\text{max}}]$ are nested, and their lengths go to zero as $h \to \infty$, by the Nested Interval Theorem, there exists a unique limit point:
\begin{equation}
    \lim_{h\to \infty} k_{\text{min}}^h = \lim_{h\to \infty} k_{\text{min}}^h = k^*.
\end{equation}
Moreover, the midpoint $k_{\text{mid}} = \frac{k_{\text{max}} - k_{\text{min}}}{2}$ also converges to $k^*$. In other words, $k_{\text{mid}} \to k^*$ as $h \to \infty$.

Finally, because $R_W(k)$ is continuous at $k^*$,  for any $\epsilon$, there exists a $\delta>0$ such that
\begin{equation}
    |k - k^*|< \delta \Rightarrow |R_W(k) - R_W(k^*)| <  \epsilon.
\end{equation}
Since $k_{\text{mid}}^h \to k^*$, for large enough $h$, $|k_{\text{mid}}^h - k^*|<\delta$. Thus, 
\begin{equation}
    |R_W(k_{\text{mid}}^h) - r| = |R_W(k_{\text{mid}}^h) - R_W(k^*)| < \epsilon.
\end{equation}
In other words, by choosing $h$ large enough (i.e., enough binary search steps), we ensure the output $k_{\text{mid}}^h$ satisfies an required error bound $|R_W(k_{\text{mid}}^h) - r_W| < \epsilon$ for $R_W$.

Finally, parameter $\lambda$ can be solved using equation $\lambda = \sqrt[m]{E(X^m)/\Gamma(1 + m/k)}$.

\end{proof}

\section{General Form Moment-based Estimator of Gamma Distribution}

\begin{thm}\label{thm:gamma}
For a Gamma random variable with the specified moments of orders $n$ and $m$, i.e., $E(X^n)$ and $E(X^m)$, where $n>m$, its parameters $\alpha$ and 
$\beta$ can be solved numerically by Algorithm \ref{alg:gamma}.
\end{thm}

\begin{algorithm}[H]
\caption{Solving Gamma parameters numerically given a pair of moments}\label{alg:gamma}
\hspace*{\algorithmicindent} \textbf{Input:} $n$, $m$, $E(X^n)$, $E(X^m)$, tolerance $\delta$\\
\hspace*{\algorithmicindent} \textbf{Output: $\alpha$, $\beta$}
\begin{algorithmic}[1]
\State $\alpha_{\text{min}} \gets \widecheck{A}$, $\alpha_{\text{max}} \gets \widehat{A}$ \Comment{choose $\widecheck{A}$ and $\widehat{A}$ such that the solution of $\alpha$ is within $[\widecheck{A}, \widehat{A}]$}
\State $r_G \gets {E^{m}(X^n)}/{E^{n}(X^m)}$
\While{$\alpha_{\text{max}} - \alpha_{\text{min}} > \delta$}
\State $\alpha_{\text{mid}} \gets {(\alpha_{\text{max}} + \alpha_{\text{min}})}/{2}$
\If{${\Gamma^{m}(1 + {n}/{\alpha_{\text{mid}}})}/{\Gamma^{n}(1 + {m}/{\alpha_{\text{mid}}})} > r_G$}
    \State $\alpha_{\text{min}} \gets \alpha_{\text{mid}}$
\Else
    \State $\alpha_{\text{max}} \gets \alpha_{\text{mid}}$
\EndIf
\EndWhile
\State $\alpha \gets {(\alpha_{\text{max}} + \alpha_{\text{min}})}/{2}$
\State $\beta \gets \sqrt[m]{E(X^m)\Gamma(\alpha)/\Gamma(m+\alpha)}$
\end{algorithmic}
\end{algorithm}

\begin{proof}
Denote the ratio of $E^{m}(X^n)$ and $E^{n}(X^m)$ as $r$, we have
\begin{equation}
    r_G = R_G(x) = \left( \frac{\beta^n \Gamma(n + \alpha)}{\Gamma(\alpha)}\right)^m \left( \frac{\beta^m \Gamma(m + \alpha)}{\Gamma(\alpha)}\right)^{-n} = \frac{\Gamma^m(n+\alpha)\Gamma^n(\alpha)}{\Gamma^n(m+\alpha)\Gamma^m(\alpha)}.
\end{equation}
In Section 4.2 of \cite{liu2025moment}, it is shown that $R_G(\alpha)$ is monotonically decreasing with respect to $\alpha$.

By the Intermediate Value Theorem (continuity), since $r \in [R_G(\alpha_{\text{min}}), R_G(\alpha_{\text{max}})]$ there exists $\alpha^{*} \in [\alpha_{\text{min}}, \alpha_{\text{max}}]$ such that $r_G = R_G(\alpha^{*})$.

At each step of Algorithm \ref{alg:gamma}, the length of the interval halves. Thus, after $h$ steps, $\alpha^h_{\text{max}} - \alpha^h_{\text{min}} = \frac{1}{2^h}(\alpha_{\text{max}} - \alpha_{\text{min}})$. In other words, the length of the interval shrinks exponentially fast.

Since the intervals $[\alpha^h_{\text{min}}, \alpha^h_{\text{max}}]$ are nested, and their lengths go to zero as $h \to \infty$, by the Nested Interval Theorem, there exists a unique limit point:
\begin{equation}
    \lim_{h\to \infty} \alpha_{\text{min}}^h = \lim_{h\to \infty} \alpha_{\text{min}}^h = \alpha^*.
\end{equation}
Moreover, the midpoint $\alpha_{\text{mid}} = \frac{\alpha_{\text{max}} - \alpha_{\text{min}}}{2}$ also converges to $\alpha^*$. In other words, $\alpha_{\text{mid}} \to \alpha^*$ as $h \to \infty$.

Finally, because $R_G(\alpha)$ is continuous at $\alpha^*$,  for any $\epsilon$, there exists a $\delta>0$ such that
\begin{equation}
    |\alpha - \alpha^*|< \delta \Rightarrow |R_G(\alpha) - R_G(\alpha^*)| <  \epsilon.
\end{equation}
Since $\alpha_{\text{mid}}^h \to \alpha^*$, for large enough $h$, $|\alpha_{\text{mid}}^h - \alpha^*|<\delta$. Thus, 
\begin{equation}
    |R_G(\alpha_{\text{mid}}^h) - r_G| = |R_G(\alpha_{\text{mid}}^h) - R_G(\alpha^*)| < \epsilon.
\end{equation}
In other words, by choosing $h$ large enough (i.e., enough binary search steps), we ensure the output $\alpha_{\text{mid}}^h$ satisfies an error bound $|R_G(\alpha_{\text{mid}}^h) - r_G| < \epsilon$ for $R_G$.

Finally, parameter $\beta$ can be solved using equation $\beta = \sqrt[m]{E(X^m)\Gamma(\alpha)/\Gamma(m+\alpha)}$.
\end{proof}

\section{General Form Moment-based Estimator of Log-normal Distribution}

\begin{thm}\label{thm:logn}
For a Log-normal random variable with the specified moments of orders $n$ and $m$, i.e., $E(X^n)$ and $E(X^m)$, where $n>m$, its parameters $\sigma$ and 
$\mu$ can be solved numerically by Algorithm \ref{alg:logn}.
\end{thm}

\begin{algorithm}[H]
\caption{Solving Log-normal parameters numerically given a pair of moments}\label{alg:logn}
\hspace*{\algorithmicindent} \textbf{Input:} $n$, $m$, $E(X^n)$, $E(X^m)$, tolerance $\delta$\\
\hspace*{\algorithmicindent} \textbf{Output: $\sigma$, $\mu$}
\begin{algorithmic}[1]
\State $\sigma_{\text{min}} \gets \widecheck{\Sigma}$, $\sigma_{\text{max}} \gets \widehat{\Sigma}$ \Comment{choose $\widecheck{\Sigma}$ and $\widehat{\Sigma}$ such that the solution of $\sigma$ is within $[\widecheck{\Sigma}, \widehat{\Sigma}]$}
\State $g \gets {E^{m}(X^n)}/{E^{n}(X^m)}$
\While{$\sigma_{\text{max}} - \sigma_{\text{min}} > \delta$}
\State $\sigma_{\text{mid}} \gets {(\sigma_{\text{max}} + \sigma_{\text{min}})}/{2}$
\If{${\Gamma^{m}(1 + {n}/{\sigma_{\text{mid}}})}/{\Gamma^{n}(1 + {m}/{\sigma_{\text{mid}}})} > g$}
    \State $\sigma_{\text{max}} \gets \sigma_{\text{mid}}$
\Else
    \State $\sigma_{\text{min}} \gets \sigma_{\text{mid}}$
\EndIf
\EndWhile
\State $\sigma \gets {(\sigma_{\text{max}} + \sigma_{\text{min}})}/{2}$
\State $\mu \gets \frac{1}{m}\log(X^m) - \frac{1}{2}\sigma^2 m$
\end{algorithmic}
\end{algorithm}

\begin{proof}
Denote the ratio of $E^{m}(X^n)$ and $E^{n}(X^m)$ as $r$, we have
\begin{equation}
    G = \frac{E(X^n)^{\frac{1}{n}}}{E(X^m)^{\frac{1}{m}}} = e^{\frac{1}{2}\sigma^2(n-m)}.
\end{equation}
In Section 4.3 of \cite{liu2025moment}, it is shown that $G$ is a function of $\sigma$, and $G(\sigma)$ is monotonically increasing with respect to $\sigma$.

By the Intermediate Value Theorem (continuity), since $r \in [G(\sigma_{\text{min}}), G(\sigma_{\text{max}})]$ there exists $\sigma^{*} \in [\sigma_{\text{min}}, \sigma_{\text{max}}]$ such that $r_G = G(\sigma^{*})$.

At each step of Algorithm \ref{alg:gamma}, the length of the interval halves. Thus, after $h$ steps, $\sigma^h_{\text{max}} - \sigma^h_{\text{min}} = \frac{1}{2^h}(\sigma_{\text{max}} - \sigma_{\text{min}})$. In other words, the length of the interval shrinks exponentially fast.

Since the intervals $[\sigma^h_{\text{min}}, \sigma^h_{\text{max}}]$ are nested, and their lengths go to zero as $h \to \infty$, by the Nested Interval Theorem, there exists a unique limit point:
\begin{equation}
    \lim_{h\to \infty} \sigma_{\text{min}}^h = \lim_{h\to \infty} \sigma_{\text{min}}^h = \sigma^*.
\end{equation}
Moreover, the midpoint $\sigma_{\text{mid}} = \frac{\sigma_{\text{max}} - \sigma_{\text{min}}}{2}$ also converges to $\sigma^*$. In other words, $\sigma_{\text{mid}} \to \sigma^*$ as $h \to \infty$.

Finally, because $G(\sigma)$ is continuous at $\sigma^*$,  for any $\epsilon$, there exists a $\delta>0$ such that
\begin{equation}
    |\sigma - \sigma^*|< \delta \Rightarrow |G(\sigma) - G(\sigma^*)| <  \epsilon.
\end{equation}
Since $\sigma_{\text{mid}}^h \to \sigma^*$, for large enough $h$, $|\sigma_{\text{mid}}^h - \sigma^*|<\delta$. Thus, 
\begin{equation}
    |G(\sigma_{\text{mid}}^h) - g| = |G(\sigma_{\text{mid}}^h) - G(\sigma^*)| < \epsilon.
\end{equation}
In other words, by choosing $h$ large enough (i.e., enough binary search steps), we ensure the output $\sigma_{\text{mid}}^h$ satisfies an error bound $|G(\sigma_{\text{mid}}^h) - g| < \epsilon$ for $G$.

Finally, parameter $\mu$ can be solved using equation $\mu = \frac{1}{m}\log(X^m) - \frac{1}{2}\sigma^2 m$.
\end{proof}

\section{Conclusion and Future Work}
In this work, we introduced a general-form moment-based estimator for the Weibull, Gamma, and Log-normal distributions that allows practitioners to estimate distribution parameters using any pair of statistical moments. We provided tailored algorithms for each distribution and proved their convergence, uniqueness, and accuracy guarantees under mild assumptions. This flexible framework unifies and generalizes many classical moment-based estimators and offers a powerful tool for modeling positively skewed, non-negative data in reliability analysis, queueing systems, production system engineering, and other applied fields.

Future work will explore extensions of this framework in several directions. First, we plan to evaluate its empirical performance across various domains and compare it quantitatively with MLE and Bayesian methods. Second, the framework may be extended to support multivariate distributions or to handle censored or incomplete data, which frequently arise in survival and reliability studies. Finally, the theoretical insights gained here could also motivate new approaches for moment-based regularization in modern machine learning applications.

\bibliographystyle{unsrtnat}
\bibliography{references}  

\end{document}